\documentclass[letterpaper,titlepage,11pt]{article}
\usepackage{amssymb,amsmath,amsfonts,enumerate}
\usepackage{graphicx}
\usepackage{subcaption}

\usepackage{boxedminipage}
\usepackage{bm}
\usepackage{float}

\def\clock{{\count0=\time
           \divide\count0 60
           \ifnum\count0<10 0\fi\the\count0
           \multiply\count0 -60 \advance\count0 \time
           :\ifnum\count0<10 0\fi \the\count0
         }}
\newcommand{\timestamp}{{\small\vbox{\hbox{\tt\jobname.tex}
\hbox{\the\day/\the\month/\the\year, \clock}}}}

\newcommand{\beq}{\begin{equation}}
\newcommand{\eeq}{\end{equation}}
\newcommand{\ben}{\begin{displaymath}}
\newcommand{\een}{\end{displaymath}}
\newcommand{\beqa}{\begin{eqnarray}}
\newcommand{\eeqa}{\end{eqnarray}}
\newcommand{\bea}{\begin{eqnarray}}
\newcommand{\eea}{\end{eqnarray}}
\newcommand{\bean}{\begin{eqnarray*}}
\newcommand{\eean}{\end{eqnarray*}}
\newcommand{\ba}{\begin{array}}
\newcommand{\ea}{\end{array}}
\newcommand{\bi}{\begin{itemize}}
\newcommand{\ei}{\end{itemize}}

\numberwithin{equation}{section}

\begin{document}

\begin{titlepage}
\begin{flushright}
\end{flushright}
\vskip 2.cm
\begin{center}
{\bf\LARGE{A Black Hole inside Dark Matter and the Rotation Curves of Galaxies}}
\vskip 1.5cm

{\bf Fateen Haddad$^1$ and Nidal Haddad$^2$
}
\vskip 0.5cm
\medskip
\textit{$^2$Department of Physics, Bethlehem University}\\
\textit{P.O.Box 9, Bethlehem, Palestine}\\

\vskip .2 in
\texttt{ f.a.assaleh@gmail.com, nhaddad@bethlehem.edu}

\end{center}

\vskip 0.3in

\baselineskip 16pt
\date{}

\begin{center} {\bf Abstract} \end{center} 

\vskip 0.2cm 

\noindent 

In this article we find a four-dimensional metric for a large black hole immersed in dark matter. Specifically, we look for and find a static spherically symmetric black hole solution to the Einstein equations which gives, in the Newtonian limit, the rotation curves of galaxies, including the flat region and the Baryonic Tully-Fisher relation, and which has a regular horizon. We obtain as well the energy-momentum tensor of the dark matter sourcing this space-time and it turns, in special, to satisfy the four energy conditions (dominant, weak, null, and strong) everywhere outside the horizon. This black-hole-dark-matter system represents a successful simplified model for galaxies, opens a new area for exploring the relativistic regime of dark matter, and shows that the theory of General Relativity together with dark matter can account for the rotation curves of galaxies.
\end{titlepage} \vfill\eject

\setcounter{equation}{0}

\pagestyle{empty}
\small
\normalsize
\pagestyle{plain}
\setcounter{page}{1}

\newpage

\section{Introduction}

Since its discovery the rotation curve of galaxies has been a fuel to research in the fields of gravitation, astrophysics, cosmology, and to intense searches for new kinds of matter. The rotation curve problem can be stated as follows: On the one hand, according to Newtonian gravity the velocity of stars around the centers of galaxies must decrease with distance (as $v\propto 1/\sqrt{r}$), but observations, on the other hand, tell that the velocities at large distances are almost constant. The rotation curve problem split the physics community into two groups. The largest group took the opinion that the solution to this problem is that most of the mass inside galaxies is not of ordinary matter (i.e., baryonic matter) but of another kind of matter which we can not observe, as it does not interact with  the baryonic matter and light except through the gravitational interaction and hence it is called "dark matter" (see for example the recent reviews \cite{Peebles:2017bzw,Bertone:2016nfn}). The second and smaller group took the opinion that it is the gravitational theories (Newtonian gravity or Einstein gravity) which must be modified in order to solve this problem (see for instance \cite{Mannheim:2012qw,Moffat:2005si,Milgrom:2019cle} and the references therein). However, this is not the whole story. After a lot of scientific research and observations another important piece of observational data, concerning the rotation curves of galaxies, showed up, the so-called (baryonic) Tully-Fisher relation \cite{Tully:1977fu,McGaugh:2000sr,Milgrom:2001ny,Milgrom:1983zz}. This is a universal relation in galaxies between the total baryonic mass and the constant velocity in the flat region. This important relation tells that the baryonic mass of a galaxy is proportional to the constant velocity to the power 4, namely, $M \propto v^4$ (see for example \cite{McGaugh:2011ac}). Thus, any candidate theory for solving the rotation curve problem must include the following main features: (1) a constant or almost constant velocity at large distances from the core, (2) the baryonic Tully-Fisher relation, and there are, in fact, two more natural requirements, (3) the Keplerian rotation curve must appear in some small region before the onset of the flat region, and (4) the theory must be contained in a relativistic framework.

In the first part of this work we reproduce some known results in the literature: we make an analysis in the non-relativistic regime and show that under the assumption that most of the baryonic mass in galaxies is concentrated in the cores, if the gravitational force inside galaxies gives both the baryonic Tully-Fisher relation and the flat part of the rotation curve, then the force must have the single form $F=-GM_bm/r^2-ma_0$ where $M_b$ is the baryonic mass of the galaxy and $a_0>0$ is a constant acceleration with a universal value. This conclusion, however, does not depend on whether we are adopting the dark matter or the modified-gravity theories. Afterwards, as an example, we show how this new force gives the main features of the observed rotation curve of the Milky Way to a very good precision. 

In the second part of this work we introduce our new results. We assume that the correct relativistic framework is the Einstein field equations and we adopt the dark matter solution to the rotation curve problem. Since very few information is known about dark matter and its nature, we found it interesting to make a simplified model of galaxies which could make us learn something new about dark matter. In this simplified  model we assume that all the baryonic mass of galaxies is enclosed inside a black hole in the centre of the galaxies. In more details, we look for and find a black hole solution to the Einstein equations which gives in the Newtonian limit (namely, the non-relativistic and weak field limit) the rotation curve of galaxies, including the flat part and the baryonic Tully-Fisher relation, and which, furthermore, has a regular horizon. The black hole solution that we find is a perturbative solution based on a derivative expansion method, which is motivated by the fact that the gravitational effects of dark matter in galaxies appear at large distances from the centers (several kilo-parsecs) and by the fact that changes due to dark matter occurs on large scales (kilo-parsecs) as well. We compute the dark matter energy-momentum tensor sourcing this black hole space-time, and find that it satisfies the four well-known energy conditions (the dominant, the weak, the null, and the strong energy conditions) showing that the dark matter obtained in this work is a manifestly physical matter field. In addition, we find that the dark matter has a negligible pressure far away from the black hole -  which is also expected - but interestingly near the black hole we find that the pressure is significant and of the same order as the mass-energy density, which signals a new "place" for learning about dark matter in the relativistic regime. The success of this black-hole-dark-matter system in producing the rotation curves of galaxies gives a strong indication that the theory of General Relativity can account for the rotation curves of galaxies without being modified.

We organise the article as follows. Sec.[\ref{sec:gra}] contains the non-relativistic preliminary analysis. We write down the force which gives both the baryonic Tully-fisher relation and the flat part of the rotation curve, we determine the value of the universal constant acceleration, and we give some plots showing how this force captures the main features of the rotation curve of the Milky Way. Sec.[\ref{sec:BH}] contains the relativistic analysis and the main results. We define the derivative expansion method, we find the black hole metric up to first order in derivatives, and we calculate the energy-momentum tensor of the dark matter. In Sec.[\ref{sec:disc}] we summarise the results of this work and discuss some important points.

\section{The Gravitational Force inside Galaxies}
\label{sec:gra}
In this section we reproduce some known results in the literature (documented, for example, in references \cite{Mannheim:1988dj,OBrien:2017bwr}, but with a different aim and perspective than ours) which are considered preliminary for the next main section. Here, we are going to focus on the region outside the cores of galaxies where we are assuming that all the baryonic mass is concentrated. Whether we are taking the dark matter approach for the rotation curve problem \cite{Peebles:2017bzw,Bertone:2016nfn} or the other approaches of modified gravities (see for example \cite{Mannheim:2012qw,Moffat:2005si}), in any case, in the non-relativistic regime one can write down the gravitational force inside galaxies as\footnote{Stars in galaxies are slowly moving bodies compared to light. }  

\beq\label{force}
F=-\frac{GM_bm}{r^2}-mg(r)
\eeq  
where the first term is the Newtonian force due to the baryonic matter, $M_b$ is the baryonic mass of the galaxy, $m$ is the mass of a test particle (e.g., a star), and the extra force $-mg(r)$ is the dark matter contribution to the gravitational force in  theories of dark matter, or it is the modification to the Newtonian force in modified-gravity theories.  In what follows we are going to determine the function $g(r)$ from the requirement that the gravitational force gives a flat, or almost flat, rotation curve at large distances and also the baryonic Tully-Fisher relation. 

At small distances from the core the Newtonian force due to baryonic matter dominates
\beq
F\approx-\frac{GM_bm}{r^2}
\eeq  
and from the circular motion equation, $a=v^2/r$, one obtains the famous Keplerian result that the circular velocity is
\beq
v=\sqrt{\frac{GM_b}{r}}
\eeq  

At large distances from the core the extra force dominates 
\beq
F\approx-mg(r)
\eeq  
and since this force must be attractive  (to cancel the centrifugal force) we conclude that $g(r)$ must be positive. Here, from $a=v^2/r$ the velocity of circular orbits is
\beq\label{flat}
v=\sqrt{g(r)r}
\eeq
However, we are not going to rush and ask that "$g(r)r=\textrm{constant}$" since it might also be possible that $g(r)r$ is only almost constant on galactic scales, that is, $g(r)r$ might be a very slowly varying function of $r$ on galactic scales. Therefore we find it instructive to impose first the baryonic Tully-Fisher relation and thereafter come back to this point.

\subsection{Imposing the Baryonic Tully-Fisher Relation}

The baryonic Tully-Fisher relation is a relation between the baryonic mass of galaxies and the constant velocity in the flat part of the rotation curve \cite{Tully:1977fu,McGaugh:2000sr,Milgrom:2001ny,Milgrom:1983zz}. The  relation reads,
\beq
M_b \propto v^4_f
\eeq
where $M_b$ is the baryonic mass of the galaxy and $v_f$ is the velocity in the flat part. Now, it is easy to see that the flat part of the rotation curve starts approximately where the two forces in Eq.[\ref{force}] are equal (or slightly after that), that is, where

\beq
\frac{GM_b}{r^2}\approx g(r)
\eeq  
 since for larger distances the baryonic Newtonian force falls off rapidly while the second force $-mg(r)$ (the force responsible for the flat part) is expected to dominate.  Let us denote the radius where the flat region starts by $r_f$ and so from the previous equation we see that this radius satisfies the following equation:
 
 \beq\label{r_f}
 g(r_f) r_f^2\approx GM_b
 \eeq
This equation, if solved, would give $r_f$ as a function of $M_b$, namely,\footnote{Note that the radius where the flat part approximately starts, $r_f$, must indeed depend on the baryonic mass $M_b$, because if, for example, the mass $M_b$ were increased then it would take the force $GM_bm/r^2$ more distance to become comparable with $-mg(r)$, which means that $r_f$ would increase.}

 \beq
r_f=r_f(M_b)
 \eeq
Now, as we increase the distance $r$, moving deep into the flat region,  Eq.[\ref{flat}] and Eq.[\ref{r_f}] give  

\beq
v_f^4=(g(r)r)^2\approx (g(r_f)r_f)^2\approx GM_bg(r_f)
\eeq
where, of course, we are assuming that $g(r)r$ is almost constant on galactic scales. Thus, in order to satisfy the baryonic Tully-Fisher relation, $v_f^4\propto M_b$, we clearly, must have 
\beq\label{a-zero}
g(r)=\textrm{constant}\equiv a_0
\eeq
because otherwise we will have $g(r_f)=g(M_b)$  (since $r_f=r_f(M_b)$) and the baryonic Tully-Fisher relation will not be satisfied. 

In summary, we have two interesting results.  First, the baryonic Tully-Fisher relation reads
\beq
M_b \approx\frac{v^4_f}{Ga_0}
\eeq
where we have determined the universal proportionality factor in terms of $a_0$; this implies, in particular, that $a_0$ itself is a universal constant. Second, we have reached the important conclusion that the gravitational force inside galaxies - and outside the cores - is given by
\beq\label{force2}
F=-\frac{GM_bm}{r^2}-ma_0
\eeq  
where $a_0>0$ is a constant acceleration. This acceleration has already been noticed in the data and analysis of rotation curves \cite{Mannheim:1988dj,OBrien:2017bwr,McGaugh:2016leg}\footnote{ Our $a_0$ is the constant called $c^2\gamma/2$ in \cite{Mannheim:1988dj} or $c^2\gamma_0/2$ in \cite{OBrien:2017bwr} and it is not the $a_0$ of MOND \cite{Milgrom:2001ny}.}. According to the references just mentioned this is the acceleration below which the Newtonian force due to the baryonic mass must be replaced with the dark matter contribution or by a modified force. Note that this is exactly what happens here: The radius where the flat region starts occurs where 
\beq\label{rf}
\frac{GM_b}{r^2}\approx g(r)=a_0
\eeq 
and this splits the galaxy into two parts, one part with $\frac{GM_b}{r^2}>a_0$ where the baryonic force prevails, corresponding to small distances, and a second part with $\frac{GM_b}{r^2}<a_0$ where the baryonic force becomes subdominant, with respect to $ma_0$, corresponding to large distances.

Before we move on to the next  section we would like to highlight an immediate and important result which comes out form this analysis. From Eq.[\ref{r_f}] it is immediately seen that
\beq\label{rf1}
r_f\approx \sqrt{\frac{GM_b}{a_0}}
\eeq 
which tells that the radius where the flat curve starts increases monotonically with the baryonic mass of the galaxy according to a square-root relation. It is worth pointing out that it is natural and expected that $r_f$ increases with $M_b$ because if the mass $M_b$ is increased then it will take the force $-GM_bm/r^2$ more distance to become comparable to $-ma_0$.

\subsection{Determining the Order of Magnitude of $a_0$ from Observational Data}
From Eq.[\ref{force2}], by assuming circular orbits, we can obtain the full rotation curve easily,
\beq\label{rot vel}
v^2=\frac{GM_b}{r}+a_0 r
\eeq  
and from this we can extract the constant (universal) acceleration  $a_0$
\beq\label{rot}
a_0=\left(v^2-\frac{GM_b}{r}\right)/r
\eeq  
By plugging in correct observational values for the triple $v$, $M_b$, and $r$  in the previous equation one can obtain $a_0$. The value of $a_0$ is supposed to be universal for (spiral) galaxies. To determine the order of magnitude of $a_0$ take a typical spiral galaxy with velocity in the flat part of the curve between $v=200\ km/s$ and $v=250\ km/s$, with baryonic mass between $M_b\approx 0.5\times 10^{11}M_{sun}$ and $M_b\approx 2\times10^{11}M_{sun}$ (see for example the review \cite{Sofue:2017}), and take a point at a distance between $r=30\ kpc$ and $r=40\ kpc$; we have chosen points in the flat part to make sure that indeed all the baryonic mass $M_b$ lies inside. With the above "wide" ranges of the parameters $v$, $M_b$ and $r$ (we have taken wide ranges deliberately to show that the order magnitude of $a_0$ is insensitive) we get that $a_0$ must lie in the range:
\beq
a_0\in [1.2 , 5.9]\times 10^{-11}\ m/s^2
\eeq  
Thus, we see that the order of magnitude of $a_0$ is definitely 
\beq
a_0\sim10^{-11}\ m/s^2
\eeq 
It is to be mentioned here that the value obtained in  \cite{OBrien:2017bwr}  ($a_0=1.4\times 10^{-11}\ m/s^2$)  lies in the above range\footnote{In reference \cite{OBrien:2017bwr} $a_0$ is called $c^2\gamma_0/2$.}. In order to obtain the exact value of $a_0$ we must do a best fitting over a large number of spiral galaxies, a thing which we do not do in this paper. In what follows we are going to make a rough estimation and determine $a_0$ from the Milky Way data. 
\subsection{The Constant Rotation Curve}

Now it is time to show that the proposed force gives indeed the flat part of the rotation curve. As an example, we take the Milky Way galaxy, and we begin by obtaining a rough estimation of the acceleration $a_0$ as follows. For a point in the flat part, say at  $r=30\ kpc$ \footnote{The radius $r=30\ kpc$ (the radius of the galaxy's disk) is considered as ideal since it encloses all the baryonic mass of the galaxy, $M_b$, whereas radii that are not very far from the core will enclose a mass that is slightly less than $M_b$.}, with reasonable values $v=200\ km/s$ and $M_b= 1\times10^{11}M_{sun}$ (see, for example, the references \cite{Sofue:2017,Sofue:2015} )\footnote{See table 6 in the review \cite{Sofue:2017}.}, we obtain
 \beq
a_0=2.8\times10^{-11}\ m/s^2
\eeq 

 If we plot Eq.[\ref{rot vel}] using the above values $a_0=2.8\times10^{-11}\ m/s^2$ and $M_b= 10^{11}M_{sun}$ we obtain the following rotation curve:

\begin{figure}[H]
\begin{center}
\includegraphics[bb=0 0 500 380 ,scale=0.5]{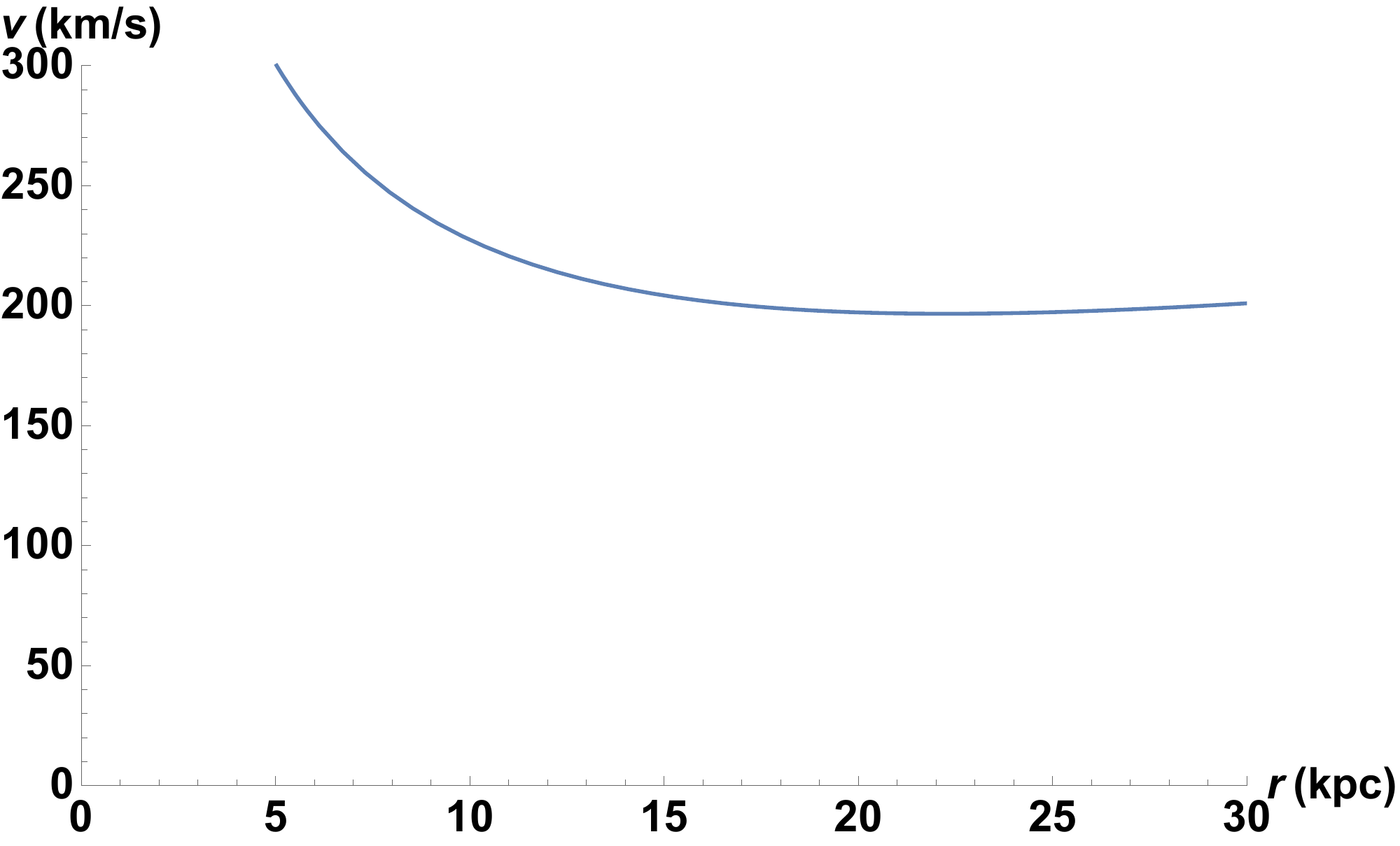}

\end{center}
\caption{\small The rotation curve of the Milky Way. The flat region is obvious in the outer half of the galaxy with a constant velocity of about $200\ km/s$. This graph must not be taken seriously in the core of the galaxy (say $r<5\ kpc$) since our treatment in this section focuses only on the region outside the core.
}\label{fig:FIG1}
\end{figure}
The flat part is clear as the velocity is almost constant at about $200\ km/s$ for several kilo-parsecs, and the Keplerian curve is also clear just before the beginning of the flat part. 
In the next plot we make a zoom-in on the flat region:

\begin{figure}[H]
\begin{center}
\includegraphics[bb=0 0 500 380 ,scale=0.5]{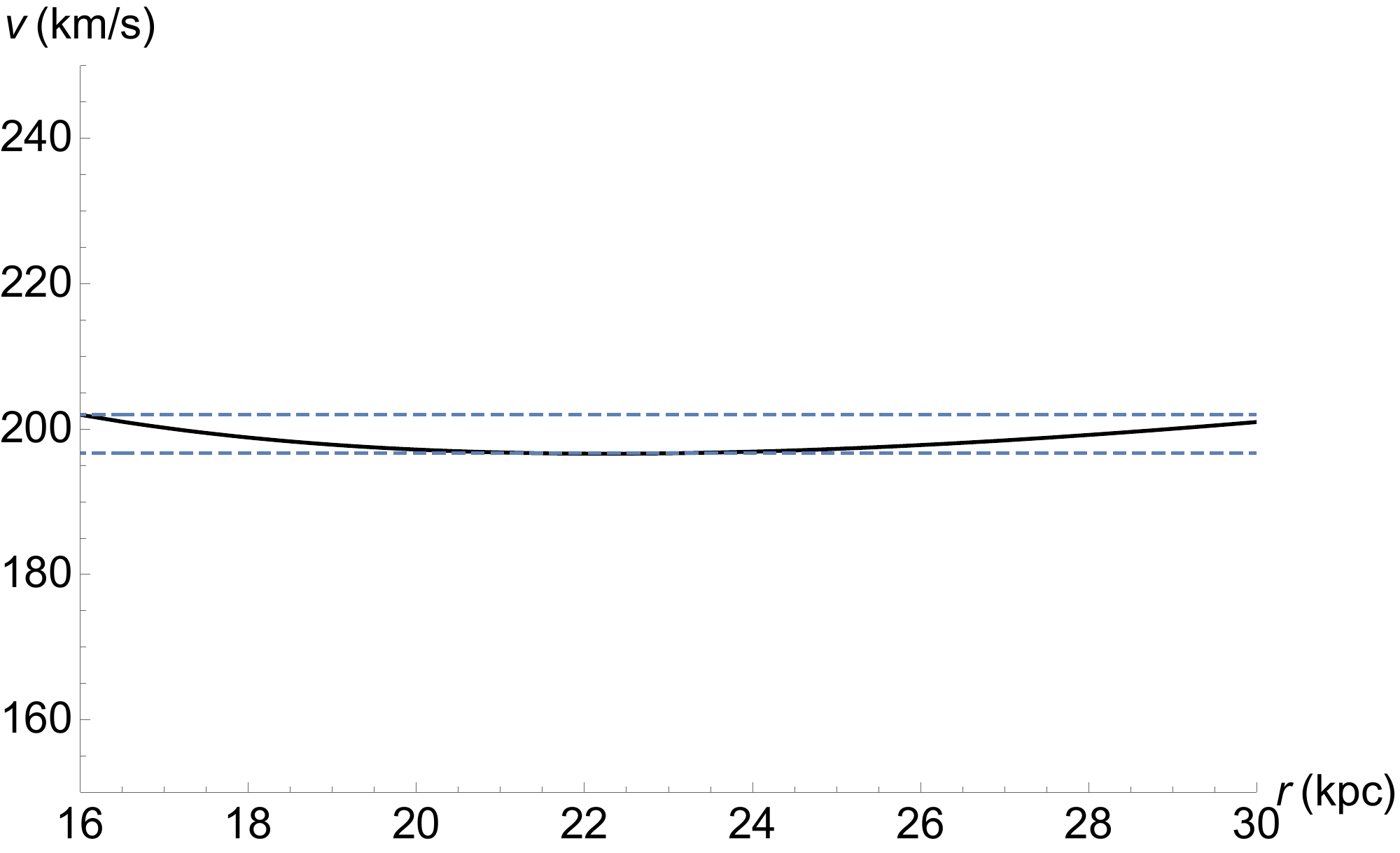}
\end{center}
\caption{\small The rotation curve of the Milky Way from 16-30\ kpc (solid line) is almost constant at about 200\ km/s. The two horizontal dashed lines $v=202\ km/s$ and $v=197\ km/s$ are the maximum and minimum velocities in this region.  
}\label{fig:FIG2}
\end{figure}

We have chosen to start at $16\ kpc$ because according to equation Eq.[\ref{rf1}] that is where the flat region starts.
Clearly the curve is almost constant along a huge distance (along almost half of the galaxy size): from $16-30\ kpc$ the velocities lie in the small range $[197\ km/s,202\ km/s]$. 

It is worth to calculate as well the velocity of the solar system about the galaxy, which we found to be  
 \beq
v_0=243\ km/s
\eeq 
and, interestingly, this is very close to values obtained in recent works (see table 2 in reference \cite{Sofue:2017}), where we have taken the distance between the solar system and the centre of the galaxy to be $R_0=8.3\ kpc$. This result shows that our model works very well not only for very far distances from the core but also for close distances down to $8\ kpc$ at the least.

As said in the beginning, our work in this section focuses only on the region outside the cores of galaxies, where most of the baryonic mass is concentrated. Therefore, the above proposed  rotation curve of the Milky Way must not be taken seriously inside the core (say for $r<5\ kpc$)\footnote{Our assumption that most of the baryonic mass in the galaxy lies within the first $5\ kpc$ has support in the literature \cite{Freeman:1970mx,Sackett:1997wf,Bovy:2013raa,Porcel:1997un} where an exponential disk is assumed, with surface mass density $M(r)\propto\exp(-r/a)$, where the scale length $a$  determines how much mass there is in the core.}, since inside the core the baryonic mass is distributed according to some distance-dependent density, $\rho=\rho(r)$, which we are not going to study. 
However, for the sake of illustration of how our work may merge with other works treating the rotation curve inside the core, let us assume a constant density of the baryonic matter inside the core, $\rho=\textrm{constant}$. Then the velocity increases linearly with distance (like a rigid body), $v\propto r$. Below we give two figures showing how the two physics, inside and outside the core, can be  approximately merged into one picture.

\begin{figure}[H]
\centering

\begin{minipage}[b]{0.45\linewidth}
\includegraphics[bb=0 0 250 380 ,scale=0.35]{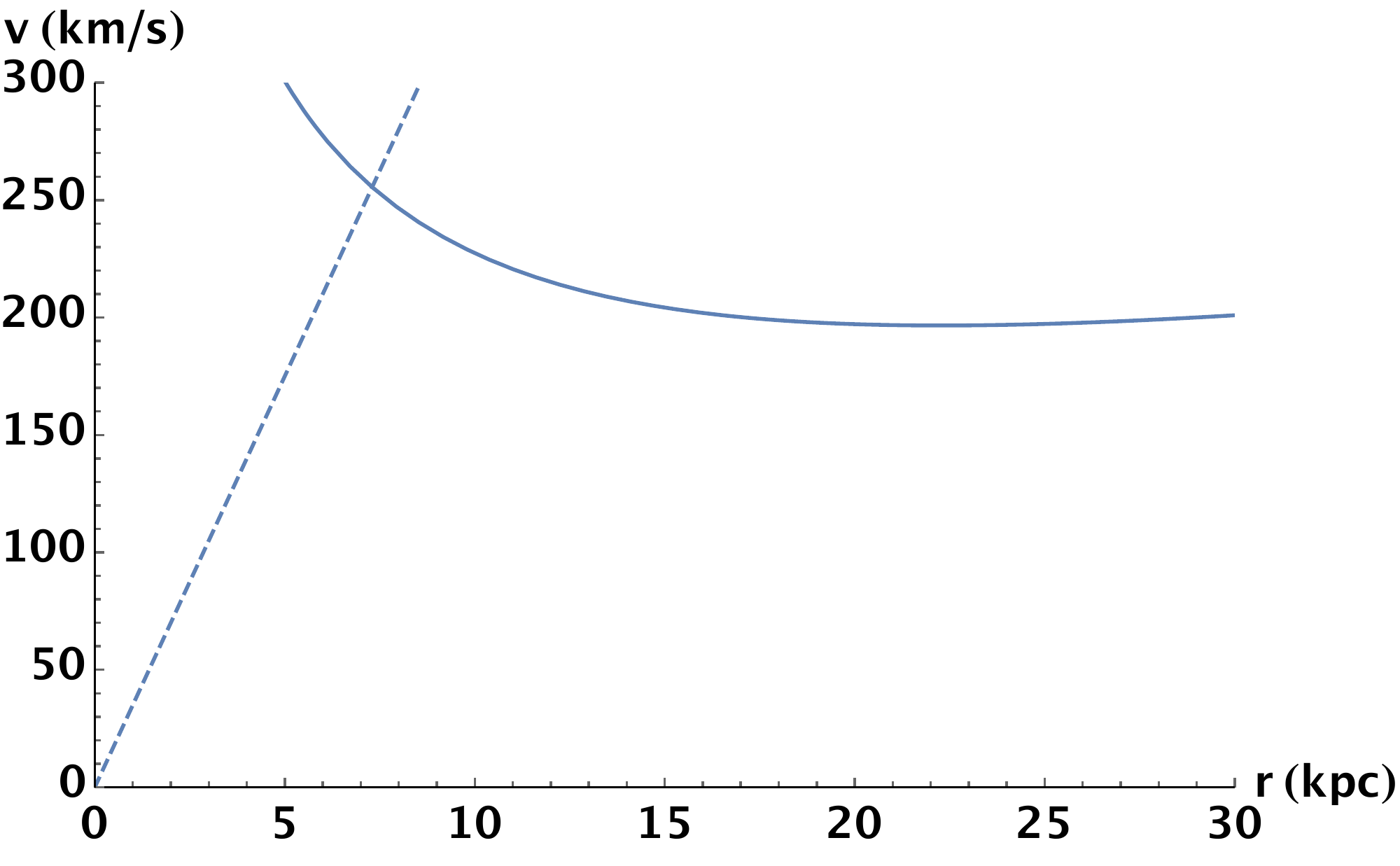}
\label{fig:minipage01}
\end{minipage}
\quad
\begin{minipage}[b]{0.45\linewidth}
\includegraphics[bb=0 0 250 380 ,scale=0.35]{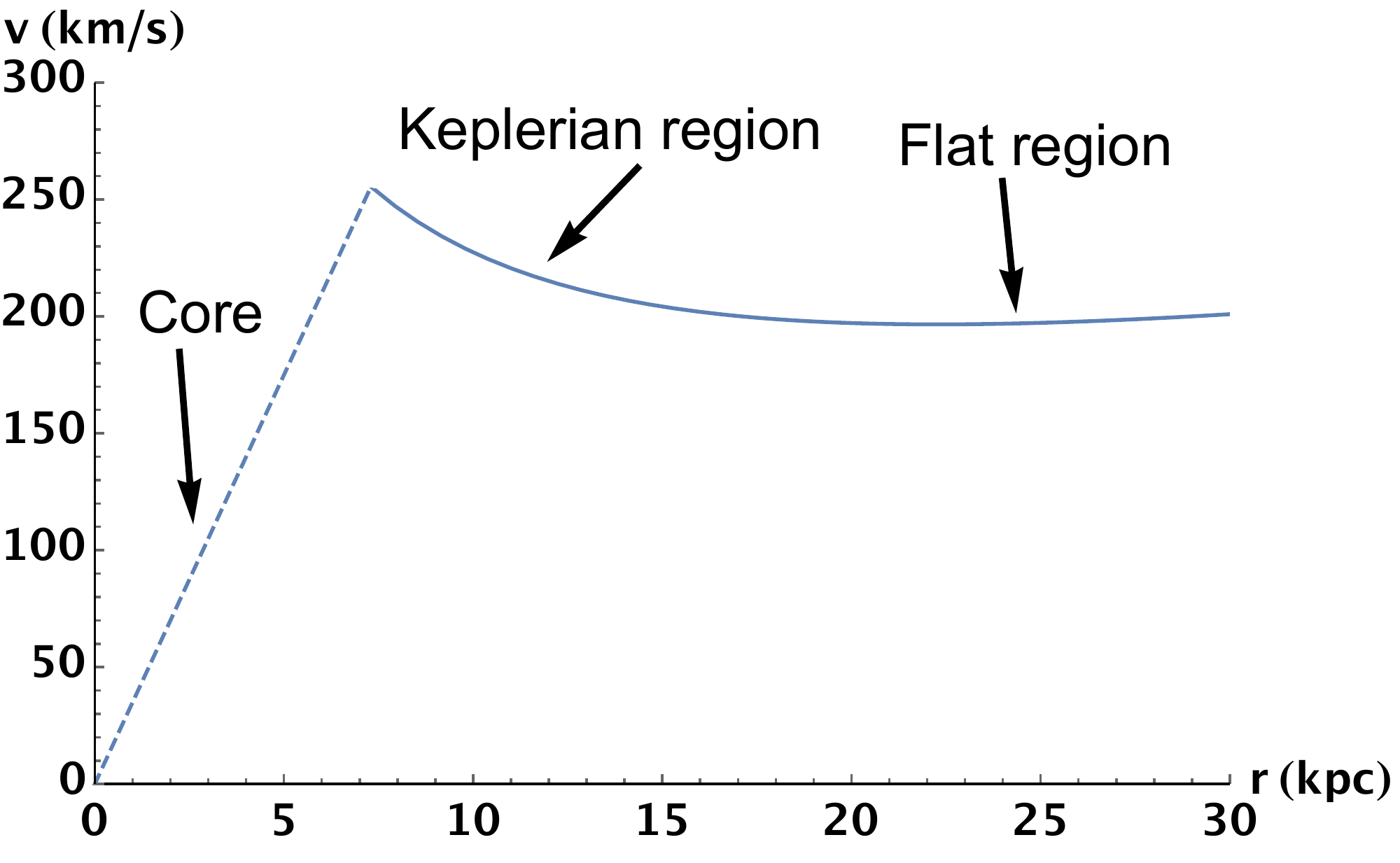}
\label{fig:minipage02}
\end{minipage}

\caption{\small Left: Our plot of the rotation curve (solid line) and the plot of the rotation curve of a rigid body (dashed line), which we took for illustration only, are put together . Right: Since our plot is valid outside the core and the linear curve (straight line) is the model we took inside, the two plots can be merged (in an approximate way) together by simply taking off the additional parts after their intersection.
 }

\label{fig:FIG0}
\end{figure}
This combined plot resembles the observed rotation curve. It contains the main features, such as, the Keplerian region, the flat region, and the peak\footnote{Many rotation curves display a peak between the core and the flat part.}. It is to be said that the way we have merged the two physics is a crude approximation; in the merging region one expects the physics to be rather complex since, for example, one can not tell exactly where the core region ends and how the density of baryonic matter changes behaviour there, a thing which could lead to a wide and unexpected merging profile as, in fact, is seen clearly in the observed rotation curve of the Milky Way.

\subsection{The Corresponding Mass Density}
For completeness and for later use we will write down here the gravitational potential which gives the force $F=-GM_bm/r^2-ma_0$ (see Eq.[\ref{force2}]) and also the corresponding mass density in galaxies. It is clear that the potential is $\phi=-GM_b/r+a_0r$. From the Poisson equation $\nabla^2\phi=4\pi G\rho$ one gets that the mass density inside galaxies is 
\beq\label{density}
\rho=\frac{1}{2\pi G}\frac{a_0}{r}
\eeq 
where as stated many times we are talking about the region outside the cores.

\section{A Black Hole inside Dark Matter}
\label{sec:BH}
In this main section we are going to look for a black hole solution to the Einstein equations which encloses all the baryonic matter of the galaxy and which gives in the Newtonian limit (the non-relativistic and weak-field limit) both, the flat rotation curve and the baryonic Tully-Fisher relation, and that, moreover, has a regular horizon.  The general static spherically symmetric metric can be written as (see for example \cite{Weinberg:1972kfs,Wald:1984rg})

\beq\label{metric}
ds^2=-f(r)dt^2+h(r)dr^2+r^2\left(d\theta^2+\sin^2\theta d\phi^2\right)
\eeq  
We find it suitable for our purposes to write the functions $f(r)$ and $h(r)$ as
\beq\label{fh}
f(r)=1-\frac{2GM_b}{r}+V(r) \qquad \qquad h(r)=\left(1-\frac{2GM_b}{r}+H(r)\right)^{-1}
\eeq  
where $M_b$ is the baryonic mass of the galaxy (enclosed inside the black hole), and where $V(r)$ and $H(r)$ are functions to be determined. 
\subsection{Derivative Expansion}
Notice that the functions $V(r)$ and $H(r)$ are the contributions of the dark matter to the metric\footnote{The functions $V(r)$ and $H(r)$ are indeed the contributions of the dark matter to the metric since if the dark matter were absent then we would simply have the Schwarzschild black hole metric.}, and therefore are expected to appear and to change significantly only on large scales (on galactic scales\footnote{From the rotation curves of galaxies we know that the effects of dark matter appear after several kilo-parsecs from the centre, and we also know that outside the cores things change significantly on the scale of kilo-parsecs as well, which is much larger than the scale of a black hole containing all the baryonic mass of the galaxy, which has a radius (scale) that is much smaller than one parsec.}). Therefore, the derivatives of $V(r)$ and $H(r)$ are expected to be very small and hence it is useful to Taylor expand them around the black hole radius $r=r_0$,
\beq\label{}
V(r)=V_0+V_1(r-r_0)+\frac{1}{2}V_2(r-r_0)^2+...
\eeq  
\beq\label{}
H(r)=H_0+H_1(r-r_0)+\frac{1}{2}H_2(r-r_0)^2+...
\eeq  
where $V_0=V(r_0)$ , $V_1=dV(r_0)/dr$, $V_2=d^2V(r_0)/dr^2$, etc., and similarly for $H(r)$. Saying that the derivatives of  $V(r)$ and $H(r)$ (namely, $V_1,H_1, V_2, H_2, etc.$ ) are very small means that  
\beq\label{}
1>>V_1(r-r_0)>>\frac{1}{2}V_2(r-r_0)^2>>...
\eeq  
\beq\label{}
1>>H_1(r-r_0)>>\frac{1}{2}H_2(r-r_0)^2>>...
\eeq  
even for $r$'s of galactic scales (in fact, this is the key point in this derivative expansion\footnote{A similar derivative expansion method for solving the Einstein equations, but in a different context, was developed in \cite{Haddad:2012ss}.}). 
\subsection{Regularity and Position of the Horizon}
The regularity of the horizon is achieved only if (see Appendix \ref{app:Edd})

\beq\label{}
h(r)=f(r)^{-1} \qquad \textrm{as}\qquad r\rightarrow r_0
\eeq 
and as can be easily seen this is satisfied only if 
\beq\label{}
H_0=V_0
\eeq 
The position of the horizon, on the other hand, is obtained from the equation $f(r_0)=0$, which gives \footnote{It is clear that the quantity $V_0$ determines the amount of dark matter inside the black hole because it  plays a role in determining its radius. Moreover, it is also clear that $V_0<0$ since the presence of dark matter is supposed to increase the mass inside the black hole and hence  its radius too.}
\beq\label{}
r_0=\frac{2GM_b}{1+V_0}
\eeq 
Since the dark matter is not expected to be significant in the cores of galaxies we will assume in this work that it is also insignificant inside our black hole, and thus we can neglect $V_0$ and set
\beq\label{}
H_0=V_0=0
\eeq 
though we expect this quantity to have interesting effects in general\footnote{The cases with  $H_0=V_0\neq 0$ will be studied in a future work.}. Thus we have 
  \beq\label{}
r_0=2GM_b
\eeq 

\subsection{Imposing the Rotation Curves of Galaxies in the Newtonian Limit}
Here we will impose the boundary condition that in the Newtonian limit (the non-relativistic and weak-field limit) the metric Eq.[\ref{metric}]  gives the rotation curve of galaxies, that is, it gives both the flat region of the rotation curve and also the baryonic Tully-Fisher relation. To do so let us assume as usual circular motions. Then, by consulting the geodesic equations one finds that (see Appendix \ref{app:geod})
\beq\label{}
\frac{v^2}{r}=\frac{1}{2}\frac{df}{dr}
\eeq 
Up to first order in derivatives one gets that

\beq\label{}
v^2=\frac{GM_b}{r}+\frac{V_1r}{2}
\eeq 
Note that this expression for $v^2$ has the same form as the one obtained in the Newtonian treatment Eq.[\ref{rot vel}] and thus we make the identification

\beq\label{}
V_1=2a_0
\eeq
 Interestingly, we have obtained the result that the rotation curve for this black hole is simply
 \beq\label{}
v^2=\frac{GM_b}{r}+a_0r
\eeq 
 and it holds for any radius from the horizon tell the far away region.
 \subsection{The Einstein Equations }
Now we are going to impose (solve) the Einstein equations
  \beq
E_{\mu\nu}=8\pi GT_{\mu\nu}
\eeq
 where $E_{\mu\nu}=R_{\mu\nu}-\frac{1}{2}Rg_{\mu\nu}$ is the Einstein tensor. We are going to assume the most general spherically symmetric energy-momentum tensor, which is that of
  an anisotropic fluid 
\beq
T_\mu^\nu=\textrm{diagonal}(-\rho,P_r,P_\perp,P_\perp)
\eeq
where $\rho$ is the mass-energy density, $P_r$ is the radial pressure, and $P_\perp$ is the transverse pressure (see references \cite{Bowers:1974tgi,Raposo:2018rjn}).
 
 One can check that the equation $E_{00}=8\pi GT_{00}$ gives (see for example \cite{Wald:1984rg})
 \beq\label{}
\rho =\frac{-h+h^2+rh'}{8\pi G r^2h^2}
\eeq
 By inserting the expression of $h(r)$ given in Eq.[\ref{fh}] in the above equation one obtains up to first order in derivatives that:
 \beq\label{}
\rho(r)=\frac{-H_1\left(r-GM_b\right)}{4\pi Gr^2}
\eeq
 To fix $H_1$ let us take the large-radius limit (the weak field limit) of $\rho$, 
 \beq\label{}
\rho(r)=\frac{-H_1 }{4\pi Gr}
\eeq
and compare it to the expression for the mass density obtained in the Newtonian treatment of the first section (see Eq.[\ref{density}])\footnote{Notice that this is another boundary condition.}. Upon comparison we see that we must fix:
\beq
H_1=-2a_0\eeq
Therefore, the mass-energy density in space up to first order in derivatives is:
 \beq\label{dens}
\rho(r)=\frac{a_0\left(r-GM_b\right)}{2\pi Gr^2}
\eeq
where note that it is positive outside the horizon.
Thus we have at hand now the metric Eq.[\ref{metric}] up to first order in derivatives 
\beq\label{metric1}
ds^2=-\left(1-\frac{2GM_b}{r}+2a_0(r-2GM_b)\right)dt^2+\frac{dr^2}{1-\frac{2GM_b}{r}-2a_0(r-2GM_b)}+r^2d\Omega^2
\eeq  
which is one of the main results of our paper\footnote{This black hole is different from the black hole given in \cite{Mannheim:1988dj} by the crucial minus sign multiplying $a_0$ in the $g_{rr}$ component. In fact, the black hole in \cite{Mannheim:1988dj} gives a negative mass-energy density if plugged into the Einstein equations.}. Here, $d\Omega^2=d\theta^2+\sin^2\theta d\phi^2$.

 The remaining of the Einstein equations ($E_{rr}=8\pi GT_{rr}$ and $E_{\theta\theta=}8\pi GT_{\theta\theta}$) will be used to obtain $P_r$ and $P_\perp$ as we show next.
From the equation $E_{rr}=8\pi GT_{rr}$ we obtain

\beq
P_r=\frac{f(1-h)+rf'}{8\pi G r^2fh}
\eeq
which up to first order in derivatives gives
\beq\label{Pr}
P_r=-\frac{a_0M_b}{2\pi r^2}
\eeq
From the equation $E_{\theta\theta}=8\pi GT_{\theta\theta}$ we obtain
\beq
P_\perp=\frac{-rhf'^2-2 f^2h'-rff'h' + 2 fh (f' + r f'')}{32\pi G rf^2h^2}
\eeq
and to first order in derivatives
\beq\label{Ppr}
P_\perp=\frac{a_0M_b}{4\pi r^2}
\eeq

For completeness, the Ricci scalar for our black hole metric is 
 \beq\label{}
R=\frac{4a_0\left(r-GM_b\right)}{r^2}
\eeq
and so we see that it is positive outside the black hole and we see again that the space-time is regular on the horizon as the Ricci scalar is finite there.

\subsection{Energy Conditions } 
Here we will show that the energy-momentum tensor found above satisfies the four energy conditions: the dominant, the weak, the null, and the strong, and hence the dark matter field obtained in this work is a manifestly physical one. For anisotropic fluids the four energy conditions are written as (see reference \cite{Hawking:1973})

\begin{align}
\rho& \geq |P_i| &\textrm{(dominant)}\\
\rho \geq 0 \qquad &\textrm{and} \qquad \rho+P_i\geq 0  &\textrm{(weak)}\\
\rho&+P_i\geq 0  &\textrm{(null)}\\
\rho+P_i\geq 0 \qquad &\textrm{and} \qquad \rho+\sum_i P_i\geq 0  &\textrm{(strong)}
\end{align}
One can easily check that the density $\rho$ and the pressures $P_r$ and $P_\perp$ given in the above section, in Eq.s [\ref{dens}],[\ref{Pr}],[\ref{Ppr}], satisfy the above four conditions\footnote{Note that the dominant energy condition implies the weak and null, while the strong implies only the null.} outside the black hole horizon (located at $r=2GM_b$)\footnote{Being more precise, the four energy conditions are satisfied for $r>GM_b$. }.

\subsection{Validity of the Derivative Expansion}
In the large-radius limit ($r>>2GM_b/c^2$) our metric [\ref{metric1}] reads 

\beq\label{metric2}
ds^2=-\left(1-\frac{2GM_b}{c^2r}+\frac{2a_0r}{c^2}\right)c^2dt^2+\left(1-\frac{2GM_b}{c^2r}-\frac{2a_0r}{c^2}\right)^{-1}dr^2+r^2d\Omega^2
\eeq  
where we have returned the speed of light $c$ to the metric. Upon recalling that $a_0\sim10^{-11}\ m/s^2$ we see that indeed
{\beq\label{}
-\frac{2GM_b}{c^2r}\pm\frac{2a_0r}{c^2}<<1
\eeq 
since the constant $a_0/c^2$ is very small in galactic scales,
 
\beq\label{}
\frac{a_0}{c^2}\sim 10^{-28}\ m^{-1}
\eeq 
 and thus even for a large distance, say $r=100\ kpc$, we have
 \beq
 \frac{2a_0r}{c^2}\sim 10^{-6}
\eeq

\section{Summary and Discussion}
\label{sec:disc}
In the preliminary first part of the article we have made a Newtonian analysis of the rotation curves of galaxies and concluded that if both the flat curve and the baryonic Tully-Fisher relation are to be obtained then the gravitational force must be $F=-GM_bm/r^2-ma_0$, where $a_0\sim10^{-11}\ m/s^2$ is a universal constant. We have also fixed the proportionality factor in the baryonic Tully-Fisher relation in terms of $a_0$, namely, $M_b \approx v^4_f/Ga_0$. As stated before, the results of this part are already known in the literature.  

In the second and main part of the article we have proposed a black-hole-dark-matter model for galaxies, with the purpose to learn new information about dark matter, especially in the relativistic regime. The black hole metric was obtained based on a derivative expansion method that is motivated and justified by the fact that the length scale of the black hole is much smaller than the galactic scale. The black hole metric which we have found up to first order in the derivative expansion is 
\beq\label{}
ds^2=-f(r)dt^2+h(r)dr^2+r^2\left(d\theta^2+\sin^2\theta d\phi^2\right)
\eeq 
with
\beq\label{}
f(r)=1-\frac{2GM_b}{r}+2a_0(r-2GM_b)\qquad \quad  h(r)=\left(1-\frac{2GM_b}{r}-2a_0(r-2GM_b)\right)^{-1}
\eeq
where $M_b$ is the baryonic mass in space and it is enclosed inside the black hole horizon and $a_0\sim10^{-11}\ m/s^2$ is the universal constant. This metric describes a static spherically symmetric black hole immersed in dark matter. This black hole gives the galactic rotation curve with many of its essential features, such as the small Keplerian region, the large flat region, and the baryonic Tully-Fisher relation. Therefore this black hole space-time can be viewed as a simplified model of galaxies.

The energy-momentum tensor of dark matter in this space-time has the form of anisotropic fluid 
\beq
T_\mu^\nu=\textrm{diagonal}(-\rho,P_r,P_\perp,P_\perp)
\eeq
with 
\beq
\rho=\frac{a_0}{2\pi Gr}-\frac{a_0M_b}{2\pi r^2} 
 \qquad \quad P_r=-\frac{a_0M_b}{2\pi r^2} \qquad\quad P_\perp=\frac{a_0M_b}{4\pi r^2}
\eeq
We have shown that this energy-momentum tensor satisfies the four energy conditions, namely, the dominant, weak, null, and strong energy conditions, outside the black hole horizon and thus the proposed dark matter is a manifestly physical matter field. Note that the radial pressure is negative, a point which poses no problem as the four energy conditions are fulfilled. Notice, furthermore, that the pressure in the far region is negligible compared to the density; if one takes the expressions for $P_r$ and $P_\perp$ given above and compute them at, say $10\ kpc$, one gets that  $P_r\sim P_\perp \sim 10^{-12}Pa$ which is even smaller than the pressure coming from the cosmological constant ($\sim 10^{-10}Pa$). Hence, this conforms with the literature that the dark matter is pressure-less and non-relativistic in the outer regions of galaxies. Nevertheless, in the near region, close to the black hole, the mass-energy density and the pressures are of the same order of magnitude which tells that  the dark matter is relativistic there. Therefore, the region close to the horizon is a place where we can learn something new about dark matter; it is a place to learn about its relativistic nature.

It is important to highlight the result that the full rotation curve for this black hole is $v^2=\frac{GM_b}{r}+a_0r$, that is, this simple formula applies for any radius from the horizon tell the far away region. This result allows us to uncover the relativistic regime of stars; since stars in orbits close to the black hole move at relativistic speeds.

As stated before, this work (the existence of the above black hole solution) gives a strong indication that the General Theory of Relativity, together with dark matter, can account for the rotation curves of galaxies, without being modified. However, the cost is that dark matter turns out to have anisotropic pressures ($P_r\neq P_\perp$), a result which may have interesting implications.

Another point we think it is important to clarify is the following. If we look at the mass-energy density $\rho$ we see that the leading term at large distances is $a_0/2\pi Gr$ which does not depend on the details of the galaxy under consideration (it is universal). At first sight one might think this is a problem, but it is not. The baryonic disk of spiral galaxies lies inside a very large dark halo, which constitutes a reservoir with respect to the relatively small and light disk. Thus, it is natural that at leading order the baryonic disk is not felt; only the background dark matter is felt. Yet, if we go for the next-to-leading order (i.e., if we ask for more precision) in $\rho$ then the baryonic mass (the details) of the galaxy appears as it should. This is analogous to the Schwarzschild metric where at far distances at leading order the Minkowski metric dominates, corresponding to the vacuum background, and only at the next order the black hole mass appears.

We find it important to mention also that the derivative expansion method we have developed here will bring out the cosmological constant term (dark energy) at second order in derivatives, which, of course, is expected and consistent with our picture and analysis since the cosmological constant effects must enter on the cosmological scale - next to the galactic scale.

Finally, it is worth mentioning that the black hole metric we have found is interesting in itself, regardless of the galactic context we have followed in this paper. That is, this derivative expansion approach for constructing the black hole metric can be used whenever a black hole is immersed in a background which changes on a large scale compared to the black hole scale.

\subsection*{Acknowledgement}
N.Haddad was supported by an internal research grant provided by Bethlehem University.

\appendix
\section{Eddington-Finkelstein Coordinates and Regularity}
\label{app:Edd}

Since we have spherical symmetry, to find the location of the horizon we look for the null $r=\textrm{constant}$ surface,
\beq\label{}
g^{\mu\nu}\partial_\mu r \partial_\nu r=0
\eeq 
which for our metric Eq.[\ref{metric}] gives

\beq\label{}
g^{rr}=h^{-1}=0
\eeq 
To make the horizon regularity manifest we move to the new coordinate $v=t+r_*$ where $dr_*/dr=\sqrt{h/f}$, upon which our metric Eq.[\ref{metric}] becomes
\beq\label{}
ds^2=-f(r)dv^2+2\sqrt{f(r)h(r)}dvdr+r^2\left(d\theta^2+\sin^2\theta d\phi^2\right)
\eeq 
and so in order to prevent the singularity of $g_{rv}$ at the horizon (as $h\rightarrow \infty$) clearly we must have
 
 \beq\label{}
f(r)=h(r)^{-1} \qquad \textrm{as}\qquad r\rightarrow r_0
\eeq

\section{Geodesic Equation and Circular Motion}
\label{app:geod}
For the general spherically symmetric static metric 
\beq\label{}
ds^2=-f(r)dt^2+h(r)dr^2+r^2\left(d\theta^2+\sin^2\theta d\phi^2\right)
\eeq  
the geodesic equations are 
\beq\label{}
\frac{d^2x^\mu}{dp^2}+\Gamma^{\mu}_{\nu\lambda}\frac{dx^\nu}{dp}\frac{dx^\lambda}{dp}=0
\eeq  
where $p$ is a parameter along the trajectory. Because of spherical symmetry the motion will take place in a plane, and without loss of generality we will take it to be in the $\theta=\pi/2$ plane.
The radial equation of motion will be (here we are going to follow reference \cite{Weinberg:1972kfs})
 \beq\label{geod1}
\frac{d^2r}{dp^2}+\frac{h'}{2h}\left(\frac{dr}{dp}\right)^2-\frac{J^2}{r^3h}+\frac{f'}{2hf^2}=0
\eeq 
 where the prime denotes derivative with respect to $r$. There are also the two familiar relations
 
 \beq\label{geod2}
\frac{dt}{dp}=\frac{1}{f(r)} \qquad \textrm{and} \qquad r^2\frac{d\phi}{dp}=J
\eeq 
where the constant $J$ is the angular momentum per unit mass. 

 On the one hand, for a circular motion, $r=\textrm{constant}$, the equation \ref{geod1} becomes
 \beq\label{geod3}
\frac{J^2}{r^3}=\frac{f'}{2f^2}
\eeq 
On the other hand, upon combining the two equations in \ref{geod2} one gets that 
 \beq\label{geod4}
J=r^2\frac{d\phi}{dt}f^{-1}
\eeq 
Finally, if we insert equation \ref{geod4} into equation \ref{geod3} we get
 \beq\label{geod5}
\frac{(v_\phi)^2}{r}=\frac{1}{2}f'
\eeq 
where $v_\phi=rd\phi/dt$ is the angular velocity.

\end{document}